# Navigating Equity and Reflexive Practices in Gigwork Design—A Journey Mapping Experience

Alicia Boyd
New York University
New York City, USA

Danielle Cummings
Department of Defense
Hawaii, USA

Angie Zhang
University of Texas at Austin
Austin, USA

## ABSTRACT

How do we create ethical and equitable experiences on global platforms? How might UX designers and developers incorporate reflexive practices–a continuous self-evaluation of one's assumptions and biases–to mitigate assumptions and workers' experience? This tutorial will explore ways to build equitable user experiences using gig work platforms as a target use case. With the rise of gig work platforms, the informal digital economy has altered how algorithmic systems manage occasional workers; its questionable assumptions have spread worldwide. Concerns over autonomy, gamification, and worker privacy and safety are amplified as these practices expand worldwide. We will practice reflexive techniques within this context by implementing an equity-focused journey-mapping experience. Journey mapping allows designers to map out the customer experience and identify potential pain points at each step that could hinder the user experience. Using a ride-sharing scenario, participants will be guided through a custom journey map highlighting equitable considerations that can facilitate responsible user experience innovation. *NOTE: The tutorial was presented at Fairness, Accountability and Transparency Conference (FAccT '24) in Rio de Janeiro.*

## 1 IMPACT

With digital platforms emerging as the newest means to save time and make money, apps that support the gig economy are generating millions of interactions between culturally and socio-economically diverse groups of people. These interactions, and the dynamics that influence them, are often overlooked in the design of these apps as more attention is devoted to the human-computer vs the human-human interaction. More importantly, because the users of these systems are so diverse, developers and stakeholders may not anticipate potential negative consequences these systems can inflict on their users [5]. This tutorial will teach participants how to take a reflexive approach—a continual self-evaluation of one's assumptions and biases—to user experience design and identify potential challenges to creating an equitable experience for their users.

## 2 INTENDED AUDIENCE(S)

This tutorial is aimed at those involved in community engagement, researchers, developers and UX designers, but is suitable for students, instructors, and industry leaders that have a vested interest in responsible innovation and development. Furthermore, this tutorial is aimed to demonstrate how to implement reflexivity in practice. Attendees should have an interest in creating and promoting equity in technology. Participants in this tutorial will have an opportunity to share thoughts, opinions and perspectives related to the presented scenario. These opportunities will enrich the tutorial, and therefore we encourage audience to communicate in ways that acknowledge and respect people's differences and different experiences.

## 3 CONTENT

In this tutorial, we will focus on use cases from the gig economy to discuss methods for creating ethical user experiences and how to center reflexivity. Many of these applications use algorithmic management, which, in the context of the gig economy, is the use of algorithms and digital platforms to manage and govern work processes, tasks, and interactions between workers and clients [11]. While algorithmic management can offer efficiency and convenience, it also raises important ethical and labor rights concerns. Balancing the power dynamics between gig workers, clients and platform operators, addressing algorithmic bias, ensuring fair compensation, and respecting workers' rights are critical considerations in this evolving landscape. The scenario we will work on represents one specific transaction (*e.g.,* rider/driver experience from ride request to rider drop-off), which is important because these types of regular interactions could understandably shape a community over a long period.

## 4 TUTORIAL TIMELINE DESCRIPTION (90 MIN)

*INTRO TO REFLEXIVITY & JOURNEY MAPPING – 20 MIN.* We will introduce the reflexivity concept through an interactive presentation filled with polls, using mentimeter[1]. Reflexivity is a continual process of self-evaluation of how the researcher's biases, values and assumptions are imparted onto the the research process [1, 9, 10]. In this tutorial, we will use reflexivity to help designers analyze actors' emotions and experiences encountered in a particular rideshare scenario. They will discuss possible outcomes as it relates to the actors' interactions on a fictitious digital platform. Our custom journey map [7] contains reflexive prompts that will allow us to challenge decisions and avoid adverse impacts resulting from one or more platform characteristics such as the following::

- **Ratings and Reviews:** Most platforms allow clients to rate and leave reviews for gig workers based on their performance. These ratings can impact gig workers' visibility to future clients and thus, availability of future gigs. Platform algorithms can also deactivate workers' accounts based on low ratings. Unfair rating systems and practices exercised by clients perpetuate power asymmetries between customers and workers [6, 12], and can directly lead to a loss of income and difficulty finding work in the gig economy [6]. Assuming the role of platform designer, participants will explore how harms may being perpetuated in rating and reputation systems and ways to mitigate those harms.
- **Surveillance and Monitoring:** Algorithmic management uses surveillance mechanisms to monitor the progress of tasks in real-time. At a minimum, this could include GPS tracking, image and timestamp tracking, and monitoring workers' interactions with clients. Such monitoring has raised concerns among gig-workers about privacy and autonomy [13]. In designing platform features that help ensure workers adhere to guidelines, is there a risk of negative impacts caused by hyper-surveillance?

---
[1] https://www.mentimeter.com

*MEET THE ACTORS – 10 MIN.* Participants will be provided with pairs of personas who will engage in the journey. Each persona will include a profile description of a fictional user of the system, however, the persona details may be based on actual user behaviors. The personas will reveal characteristics that impact the journey map. Therefore, each group will use reflexivity to infer possible emotions and impacts experienced along the journey.

*REVEAL THE SCENARIO & JOURNEY PHASES – 25 MIN.* Next we will discuss the journey map that participants will use to apply their reflexive methodology. Every journey map consists of phases or high-level steps that a user must go through to complete a task in the system. The journey map that we will provide will consist of a scenario involving two personas using a rideshare application. In this scenario, a typical journey might include the following phases: Set up account, request a ride, ride in car, arrival. Participants may choose to include additional phases or actions during the breakout portion of the tutorial.

*MINDSETS, EMOTIONS AND OPPORTUNITIES – 20 MIN.* After learning about reflexivity and being introduced to the rideshare scenario and a persona template, participants will form teams of 3-4 to explore actions, mindsets, and emotions. In this discussion exercise, teams will assess 2 personas in the rideshare scenario: a rider and a driver. They will use a provided journey map to understand the user actions taken at each phase, and will use reflexive questions to try to understand the mindset, the users' thoughts and motivations, and the emotions, the highs and lows, at each phase. Providing this additional context should reveal opportunities for mitigating some of the potential pain points, biases and other barriers to equality. ***Note***: In-person participants will use hardcopies of the journey map. Virtual participants will use Mural[2] and breakout rooms via Zoom.

*SHARE OUT INSIGHTS & KEY TAKEAWAYS – 15 MIN.* When participants come together again, each group will share the opportunities they identified, focusing on those that represent the biggest impact to the user experience. If time allows, teams will be encouraged to discuss who owns the change needed and how might those owners measure improvement.

---

[2]https://www.mural.co

## 5 THE TEAM

**Alicia Boyd** is a Postdoctoral Associate Researcher at New York University specializes in researching reflexive practices in software engineering [3] and machine learning [2, 4]. She has over a decade of experience teaching and organizing workshops, tutorials, and seminars for practitioners and researchers.

**Danielle Cummings** is a Technical Director and Instructor for the Department of Defense, where she uses and teaches software development, human-computer interaction and Lean Startup methodology. She frequently coaches teams and individuals in the use of journey maps and other similar tools for understanding the user and their needs in the course of developing solutions. Her recent research has focused on developing culturally-centric pedagogy [8] for engaging minorities in STEM learning through activism.

**Angie Zhang** is an HCI PhD student at the University of Texas at Austin's School of Information. She researches how to design human-centered technology, including how to support app-based gig workers navigating platform algorithms through intervention design and policy-related needs. She is advised by Dr. Min Kyung Lee, who has conducted fundamental work around algorithmic management and its impact on workers.

## 6 RESOURCES AND ACCESSIBILITY

More information can be found at the Navigating Equity and Reflexive Practices in Gigwork Design website. In the virtual tutorial offering, presenters will use Mural, an online whiteboard tool, which contains several accessibility features including a screen reader, color-blind accessible palettes, zooming solutions, and keyboard navigation. Presenters will highlight alternative navigation methods during the journey map introduction.